\documentclass[a4paper,11pt]{article}
\pdfoutput=1 

\usepackage{jheppub} 
\usepackage{amsmath,amssymb}
\usepackage{graphicx}
\usepackage{subcaption}
\usepackage[percent]{overpic}
\usepackage[dvipsname]{xcolor}
\usepackage{multirow}
\usepackage{bbm}

\def\be{\begin{equation}}
\def\ee{\end{equation}}

\newcommand{\CO}{\mathcal{O}}

\newcommand{\CN}{\mathcal{N}}

\newcommand{\IC}{\mathbb{C}}
\newcommand{\IP}{\mathbb{P}}

\newcommand{\IU}{\mathbb{U}}
\newcommand{\IZ}{\mathbb{Z}}
\newcommand{\IF}{\mathbb{F}}
\newcommand{\IR}{\mathbb{R}}

\newcommand{\Hom}{\mathrm{Hom}}
\newcommand{\Arg}{\mathrm{Arg}}

\newcommand{\Eq}{\mathrm{Eq}}
\renewcommand{\-}{\text{-}}
\renewcommand{\(}{\left(}
\renewcommand{\)}{\right)}

\title{On the BPS spectrum of 5d $SU(2)$ super-Yang-Mills}
\author[a]{Pietro Longhi}
\affiliation[a]{Institute for Theoretical Physics, ETH Zurich, 8093, Zurich, Switzerland}
\emailAdd{longhip@phys.ethz.ch}

\abstract{We provide a closed-form expression for the motivic Kontsevich-Soibelman invariant for M-theory in the background of the toric Calabi-Yau threefold $K_{\mathbb{F}_0}$.
This encodes the refined BPS spectrum of $SU(2)$ 5d ${\cal N}=1$ Yang-Mills theory on $S^1\times \IR^4$, corresponding to rank-zero Donaldson-Thomas invariants for $K_{\IF_0}$, anywhere on the Coulomb branch.  \\\today}

\begin{document} 

\maketitle
\flushbottom

\section{Introduction and main result}

Structures of BPS spectra reflect deep aspects of a gauge theory's dynamics, a fact that became prominent in theories with eight supercharges with the work of Seiberg and Witten on 4d $\CN=2$ Yang-Mills theory \cite{Seiberg:1994rs}.
This paper provides an exact description of the BPS spectrum of the circle uplift of that theory, 5d $\CN=1$ Yang Mills with gauge group $SU(2)$.
BPS states of five-dimensional theories with eight supercharges are also interesting from a mathematical viewpoint, because their BPS indices correspond to rank-zero Donaldson-Thomas invariants of certain Calabi-Yau threefolds. The theory considered here corresponds to the canonical bundle of the Hirzebruch surface $\IF_0$ \cite{Seiberg:1996bd, Morrison:1996xf, Douglas:1996xp}.
Enumerative invariants of geometries with compact four-cycles are generally very difficult to compute. There are currently no examples where an explicit and exhaustive description of rank-zero Donaldson-Thomas invariants is available, with the exception of vey recent developments for local $\IP^2$ based on scattering-diagram techniques \cite{Bousseau:2019ift}.

A given geometry typically has not one, but many different BPS spectra, corresponding to different regions of moduli space and being related to each other by wall-crossing \cite{Kontsevich:2008fj, Joyce:2008pc, Gaiotto:2008cd, Gaiotto:2009hg}.
A neat way to describe all possible spectra of a given geometry (or gauge theory) at once, is to compute the wall-crossing invariant of Kontsevich and Soibelman also known as \emph{(motivic) spectrum generator} or \emph{BPS monodromy} in physics.
In this paper we will take this approach, and derive an exact expression for the motivic wall-crossing invariant $\IU$ for the canonical bundle $K_{\IF_0}$. 
The standard definition of $\IU$ involves knowing in advance the \emph{full} BPS spectrum at some point in K\"ahler moduli space.
When this information is unavailable (as in our case) a different strategy is to find a choice of moduli for which central charges are maximally aligned \cite{Longhi:2016wtv}. This is the \emph{Roman locus} of \cite{Gabella:2017hpz}, where the Kontsevich-Soibelman invariant may be obtained directly via spectral (or exponential \cite{Eager:2016yxd}) networks \cite{Gaiotto:2012rg}.
Unfortunately, the conditions for existence of Roman loci are poorly understood even for 4d $\CN=2$ theories, and much less is known about the 5d case.
In this paper we introduce a new approach to computing $\IU$, based on leveraging information about BPS states at different points in moduli space studied in \cite{Closset:2019juk, Banerjee:2020moh}.
Details of this will be explained below.

Fixing a choice of K\"ahler moduli determines a unique BPS spectrum, with BPS states of charge $\gamma\in \Gamma$ characterized by a central charge $Z_\gamma \in \Hom(\Gamma, \IC)$. Here $\Gamma\simeq \IZ^4$ is a lattice of charges endowed with a skew-symmetric bilinear form $\langle \ \cdot\ ,\ \cdot\ \rangle$ which corresponds to the Dirac-Schwinger-Zwanziger pairing in the gauge theory. 
CPT symmetry implies that if the spectrum features a state with charge $\gamma$, there is a corresponding state with charge $-\gamma$.
We henceforth focus on studying the half of the spectrum with $-\pi/2\leq \Arg Z_\gamma <\pi/2$.
The charges $\gamma$ of all such states can be described as positive-integer linear combinations of four basic charges $\gamma_1,\dots,\gamma_4$, whose definition is fixed by our choice of half-plane.
For the central charge configurations considered in this note, the $\gamma_i$ correspond to the following exceptional sheaves on $K_{\IF_0}$, or fractional D-branes in IIA string theory
\be
	\begin{array}{|c|c|c|c|}
		\hline
		\gamma_1 & \gamma_2  & \gamma_3 & \gamma_4 \\
		\hline
		\ \CO(0,0) \ & \CO(1,0)  & \CO(1,1) & \CO(2,1) \\
		\hline
		D4 & \ D2_f\-\overline{D4}\  & \ D0\-D2_b\-\overline{D2}_f\-\overline{D4}\  & \ \overline{D2}_b\-{D4}\ 
		\\
		\hline
	\end{array}
\ee
where $D2_{b}$ (resp. $D2_f$) denotes a D2-brane wrapping the base (resp. fibre) $\IP^1$ in $\IF_0$ and the overline denotes the anti-brane. 
The pairing is $\langle\gamma_{i},\gamma_{i+1} \rangle = -2$ with $i \in \IZ / 4\IZ$, see \cite{Banerjee:2020moh}.

The BPS spectrum is encoded by the BPS index $\Omega(\gamma)\in \IZ$, a supersymmetry-protected quantity whose absolute value roughly coincides with the dimension of the BPS Hilbert space.\footnote{More precisely, $|\Omega|$ coincides with the dimension of the Hilbert space for the center-of-mass degrees of freedom, and only under the  \emph{no-exotics} assumption \cite{Gaiotto:2010be, DelZotto:2014bga, Chuang:2013wt}.}
The Protected Spin Character $\Omega(\gamma;y)$ is a Laurent polynomial with integer coefficients, that refines the BPS index by encoding information about the spin of BPS states. The relation between the two is $\Omega(\gamma) = \Omega(\gamma;y=-1)$.
Mathematically the BPS index should coincide with numerical rank-zero Donaldson-Thomas invariants, while the Protected Spin Character should correspond to their motivic version.

Having spelled out the necessary conventions we can state the main result. We propose that the motivic wall-crossing invariant is
\be\label{eq:U-expression}
\begin{split}
	\IU
	& = 
	\prod_{k\geq 0}^{\nearrow}\Phi( \hat Y_{\gamma_1 + k (\gamma_1+\gamma_2)}) \Phi( \hat Y_{\gamma_3+ k (\gamma_3+\gamma_4)}) \\
	& \times 
	\prod_{n\geq 1}  \Phi((-y)^{-1} \hat Y_{n(\gamma_1+\gamma_2+\gamma_3+\gamma_4)})^{-1} \Phi((-y) \hat Y_{n(\gamma_1+\gamma_2+\gamma_3+\gamma_4)})^{-2}  \Phi((-y)^3 \hat Y_{n(\gamma_1+\gamma_2+\gamma_3+\gamma_4)})^{-1}\\ 
	& \times 
	\prod_{k\geq 0}
	\Phi((-y)^{-1} \hat Y_{\gamma_1 + \gamma_2 + k (\gamma_1+\gamma_2+\gamma_3+\gamma_4)})^{-1} \Phi((-y) \hat Y_{\gamma_1 + \gamma_2 + k (\gamma_1+\gamma_2+\gamma_3+\gamma_4)})^{-1} 
	\\
	& \times
	\prod_{k\geq 0}
	\Phi((-y)^{-1} \hat Y_{\gamma_3+\gamma_4+ k (\gamma_1+\gamma_2+\gamma_3+\gamma_4)})^{-1} \Phi((-y) \hat Y_{\gamma_3+\gamma_4+ k (\gamma_1+\gamma_2+\gamma_3+\gamma_4)})^{-1} \\
	& \times
	\prod_{k\geq 0}^{\searrow}\Phi( \hat Y_{\gamma_2 + k (\gamma_1+\gamma_2)}) \Phi( \hat Y_{\gamma_4+ k (\gamma_3+\gamma_4)}) 
\end{split}
\ee
where $\Phi(\xi) = \prod_{s\geq 0}(1+ y^{2s+1}\xi)^{-1}$ is a variant of the quantum dilogarithm function,  $\hat Y_\gamma$ are quantum-torus variables obeying $\hat Y_{\gamma} \hat Y_{\gamma'} = y^{\langle\gamma,\gamma'\rangle} \hat Y_{\gamma+\gamma'}$, and $\nearrow$ (respectively $\searrow$) denotes increasing (decreasing) values of $k$ to the right.

An important clarification is now in order. 
Factorizations of $\IU$ into products of quantum dilogarithms like (\ref{eq:U-expression}) are typically associated with a certain BPS spectrum, where each factor $\Phi((-y)^m \hat Y_\gamma)$ corresponds to a BPS state of charge $\gamma$ and spin $m$, and the ordering of factors reflects the ordering of phases of $Z_\gamma$'s.
However this is \emph{not} the case here, at least not necessarily.
In fact, we did not find a point in the moduli space of $K_{\IF_0}$ where central charges have this  configuration and where the spectrum is so simple. As a consequence, the derivation is not straightforward, and will be discussed below. More precisely, factors of lines 3 \& 4 are derived directly for $k=0,1$. We provide an algorithm to compute higher $k$, and give strong evidence for the all-order expression, plus extensive checks.

An exact expression for $\IU$ allows to study the BPS spectrum anywhere in moduli space. 
For this purpose, one must factorize $\IU$ into a product of terms $\Phi((-y)^m \hat Y_\gamma)$.
Since variables $\hat Y_\gamma$ do not commute, the factorization depends on the ordering, which corresponds to that of $\Arg Z_\gamma$ (decreasing from left to right), and is in turn fixed by a generic choice of moduli.
Exponents of the factorization correspond to Laurent coefficients of the Protected Spin Characters. 
We illustrate this in Section \ref{sec:sym-point-spectrum}, by factorizing $\IU$ at a point in moduli space with fiber-base symmetry, and obtaining the corresponding refined BPS spectrum.

\section{Derivation}

We now explain how (\ref{eq:U-expression}) is derived.
Suppose we fix a generic choice of moduli, away from walls of marginal stability.
Given any angular sector $\measuredangle$ in the complex plane, we may define $\IU(\measuredangle)$ as the phase-ordered product of $\Phi((-y)^m \hat Y_\gamma)^{a_m(\gamma)}$ for any $\gamma$ whose central charge has phase within that sector. 
Here $a_m(\gamma)\in \IZ$ are coefficients of the PSC, namely $\Omega(\gamma;y) = \sum_{m\in \IZ} (-y)^m a_m(\gamma)$.
We will split the half-plane $-\pi/2 \leq \Arg Z < \pi/2$ into three sectors: the positive real line $\IR^+$, the sector with positive real an imaginary parts $\measuredangle^+$ corresponding to $0<\Arg Z <\pi/2$ and the sector $\measuredangle^-$ corresponding to $-\pi/2 \leq \Arg Z <0$.
This defines a decomposition of the wall-crossing invariant 
\be\label{eq:U-split}
	\IU = \IU(\measuredangle^+) \cdot \IU(\IR^+) \cdot \IU(\measuredangle^-)
\ee

The wall-crossing formula of Kontsevich and Soibelman asserts that $\IU(\measuredangle)$ is invariant under changes of stability conditions (i.e. central charges \cite{Douglas:2000qw, bridgeland2007stability}), as long as no \emph{BPS rays} enter or exit the sector $\measuredangle$. By a BPS ray  we mean any locus $Z_\gamma \IR^+ \subset \IC$ such that $\Omega(\gamma;y) \neq 0$.
We will take advantage of this invariance property to compute the three factors in (\ref{eq:U-split}) at different points in the moduli space of stability conditions: in particular we will compute $\IU(\measuredangle^\pm)$ for a certain configuration of central charges, and $\IU(\IR^+)$ for a different one.
All we need to ensure is that these configurations of central charges are connected by a variation of central charges that never causes a BPS ray to cross the boundaries of  sectors $\measuredangle^\pm$.

\subsection*{A path in the moduli space of stability conditions}

We consider the mirror geometry of $K_{\IF_0}$, described by a conic bundle over the algebraic curve $1 - Q_b(x+x^{-1}) + Q_f(y+y^{-1})=0$ in $\IC^*_x\times \IC^*_y$.
The four basic charges are identified with homology cycles on this curve, see \cite{Banerjee:2020moh} for a detailed description.
Central charges are determined by periods of the 1-form $\lambda = (2\pi)^{-1} \log y \, \frac{dx}{x}$, and can be evaluated numerically.
For $Q_b=-1, Q_f=2$ one finds that $\pi/2>\Arg Z_{\gamma_3} >\Arg Z_{\gamma_1} > 0$ and $\Arg Z_{\gamma_2}= \Arg Z_{\gamma_4}=-\pi/2$, and in particular\footnote{
Numerical evaluation yields $Z_{\gamma_1} \approx 2.59433\, +0.349113 i$, $Z_{\gamma_2} \approx -0.349113 i$, $Z_{\gamma_3} \approx 3.68886+1.03718 i$,  $Z_{\gamma_4} \approx -1.03718 i$. The reality condition (\ref{eq:positivity}) can also be verified by plotting the exponential network at $\vartheta=0$, where both saddles of $\gamma_1+\gamma_2$ and of $\gamma_3+\gamma_4$ appear, see Figure \ref{fig:theta-0-network}.} 
\be\label{eq:positivity}
	Z_{\gamma_1+\gamma_2}  , Z_{\gamma_3+\gamma_4} \in \IR^+\,.
\ee
The ray $\IR^+$ contains central charges of all states in the span of $\gamma_1+\gamma_2$ and $\gamma_3+\gamma_4$. 
Note that these are mutually local, i.e. $\langle \gamma_1+\gamma_2,\gamma_3+\gamma_4\rangle = 0 $ (the same holds for charges in their span), this ensures we are not on a wall of marginal stability. This is the starting point $t=0$ of the path in the moduli space of central charge configurations 
\be\label{eq:virtual-stab-path}
\begin{split}
	\tilde Z_{\gamma_1}(t) &= (1-t) Z_{\gamma_1}  + t Z_{\gamma_3}
	\\
	\tilde Z_{\gamma_3}(t) &= Z_{\gamma_3}
\end{split}
\qquad
\begin{split}
	\tilde Z_{\gamma_2}(t) &= (1-t) Z_{\gamma_2}  + t Z_{\gamma_4}
	\\
	\tilde Z_{\gamma_4}(t) &= Z_{\gamma_4}
\end{split}
\ee
The path ends at $t=1$, where 
\be\label{eq:virtual-stab}
	Z'_{\gamma_1}=Z'_{\gamma_3} \in \measuredangle^+ \qquad Z'_{\gamma_2} = Z'_{\gamma_4}  \in \measuredangle^-\,.
\ee 
The path (\ref{eq:virtual-stab-path}) has the crucial property that
\be 
	\tilde Z_{\gamma_1}(t) + \tilde Z_{\gamma_2}(t)  = (1-t) (Z_{\gamma_1} + Z_{\gamma_2}) + t (Z_{\gamma_3}+Z_{\gamma_4})  \in \IR^+
\ee
is real and positive for all $t\in [0,1]$, thanks to (\ref{eq:positivity}).
The configuration of central charges for $t=0$ is realized at a point in the moduli space of the (mirror) geometry, and we refer to it as a \emph{physical} stability condition. 
On the other hand, the configurations $\tilde Z(t)$ for $t>0$ are not necessarily realized in moduli space, and we refer to them as \emph{virtual} stability conditions. Configurations analogous to $t=0$ and $t=1$ were considered in \cite{Banerjee:2020moh} and \cite{Closset:2019juk}.

\subsection*{Computing $\IU(\measuredangle^\pm)$ at $t=1$}

The virtual stability condition for $t=1$ is especially nice because the spectrum in $\measuredangle^\pm$ has a simple structure. This was analyzed in \cite{Closset:2019juk}, whose approach we now review. For $\IU(\IR^+)$ we will adopt a different method later, by working at $t=0$. 
The BPS spectrum admits a description in terms of the representation theory of a BPS quiver with four nodes \cite{Hanany:2001py}, our conventions are adapted to  \cite{Banerjee:2020moh}. 
For the choice of half-plane considered in this note, the structure of the quiver is cyclic, with two arrows from the $i$-th node to the $i+1$-th node.
Moreover each node is labeled by a charge: quite simply, node $i$ is labeled by $\gamma_i$.
Tilting the choice of half-plane clockwise induces the following mutation sequence
\be\label{eq:mut-seq}
	\mu_+ = \mu_4\circ \mu_2\circ\mu_3\circ\mu_1
\ee
respectively on nodes $1,2,3,4$. 
The central charges with largest phase are $Z'_{\gamma_1}, Z'_{\gamma_3}$, for this reason $\mu_3\circ \mu_1$ come first. 
To see the rest of the sequence, recall that a mutation on the node with label $\gamma$ changes the labeling as follows\footnote{The structure of the quiver at each step may be recovered by recalling that $\langle\gamma,\gamma'\rangle = m > 0$ corresponds to $m$ arrows oriented from the node with charge $\gamma'$ to the node with charge $\gamma$.} \cite{Alim:2011ae, Alim:2011kw}
\be
	\gamma \to -\gamma
	\qquad
	\gamma ' \to \gamma' + [\langle \gamma',\gamma\rangle]_+ \gamma
\ee
where $[x]_+ = {\rm max}(0,x)$.
The sequence (\ref{eq:mut-seq}) produces the following labels at each step
\be
\begin{array}{c|cccc}
	\text{quiver \textbackslash \ node} & 1 & 2 & 3 & 4 \\
	\hline
	Q & \gamma_1 & \gamma_2 & \gamma_3 & \gamma_4 \\
	\mu_1 \circ Q & -\gamma_1 & \gamma_2+ 2\gamma_1 & \gamma_3 & \gamma_4 \\ 
	\mu_3\circ \mu_1 \circ Q & -\gamma_1 & \gamma_2+ 2\gamma_1 & -\gamma_3 & \gamma_4 +2\gamma_3\\
	\mu_2\circ \mu_3\circ \mu_1 \circ Q & 2\gamma_2+ 3\gamma_1 & -\gamma_2- 2\gamma_1 & -\gamma_3 & \gamma_4 +2\gamma_3\\ 
	\mu_4\circ \mu_2\circ \mu_3\circ \mu_1 \circ Q & 2\gamma_2+ 3\gamma_1 & -\gamma_2- 2\gamma_1 & 2\gamma_4 + 3 \gamma_3 & -\gamma_4 -2\gamma_3\\ 
\end{array}
\ee
where each time the mutation is on the node with highest $\Arg Z'_\gamma$ (clockwise tilting).

In particular,  the quiver $\mu_4\circ \mu_2\circ \mu_3\circ \mu_1 \circ Q$ has the same exact structure as $Q$, and the stability condition is also analogous. 
Thus tilting the half-plane further clockwise simply iterates $\mu_+$, infinitely many times.
The BPS rays encountered in this way, within sector $\measuredangle^+$, correspond to charges $\gamma_1 + k(\gamma_1+\gamma_2)$ and $\gamma_3 + k(\gamma_3+\gamma_4)$ for $k\geq 0$. Since they appear as charges of nodes, they all have PSC $\Omega(\gamma;y)= 1$ \cite{Alim:2011ae, Alim:2011kw}.
Similar considerations apply to counter-clockwise rotations of the choice of half-plane. In this case one finds BPS states with charges $\gamma_2 + k(\gamma_1+\gamma_2)$ and $\gamma_4 + k(\gamma_3+\gamma_4)$, again with $\Omega(\gamma;y)=1$.
Note that the central charges of these towers of BPS states both asymptote to $\IR^+$, but from opposite sides. 
This leads to the conclusion that there is a \emph{single} `accumulation' ray along $\IR^+$.

Since the clockwise (resp. counter-clockwise) tilting of the half-plane covers the whole angular sector $\measuredangle^+$ (resp. $\measuredangle^-$), we can write down 
$\IU(\measuredangle^\pm)$ 
\be\label{eq:BPS-rays-pm-virtual}
\begin{split}
	\IU(\measuredangle^+) &= \prod_{k\geq 0}^{\nearrow} \Phi( \hat Y_{\gamma_1 + k (\gamma_1+\gamma_2)}) \Phi( \hat Y_{\gamma_3+ k (\gamma_3+\gamma_4)}) \\
	\IU(\measuredangle^-) &= \prod_{k\geq 0}^{\searrow}\Phi( \hat Y_{\gamma_2 + k (\gamma_1+\gamma_2)}) \Phi( \hat Y_{\gamma_4+ k (\gamma_3+\gamma_4)})  \\
\end{split}
\ee
One should worry that $\Arg Z'_{\gamma_1 + k (\gamma_1+\gamma_2)} = \Arg Z'_{ \gamma_3+ k (\gamma_3+\gamma_4)}$ may cause ordering ambiguities in the above formulae. 
But since $\langle \gamma_1 + k (\gamma_1+\gamma_2),  \gamma_3+ k' (\gamma_3+\gamma_4)\rangle=0$ and $\langle \gamma_2 + k (\gamma_1+\gamma_2),  \gamma_4+ k' (\gamma_3+\gamma_4)\rangle = 0$ when $k=k'$, the corresponding factors commute.

As we move from $t=1$ to the physical stability condition $t=0$, the BPS rays in (\ref{eq:BPS-rays-pm-virtual}) begin to move within $\measuredangle^\pm$, however they never exit these sectors. This is clear because, e.g.  $\Arg Z_{\gamma_1+k (\gamma_1+\gamma_2)}(t) > \Arg Z_{\gamma_1+\gamma_2}(t)$  for all $k\geq 0$ as long as $\Arg Z_{\gamma_1}(t) > \Arg Z_{\gamma_2}(t)$ and as long as both have positive real part. 
Since $Z_{\gamma_1+\gamma_2}(t)\in \IR^+$ which separates the two sectors $\measuredangle^\pm$, it is clear that the BPS rays are confined within each sector separately.
As BPS rays move around within $\measuredangle^\pm$ they may cross each other, and generate new BPS rays by wall-crossing.
Any new rays generated in this way must lie in the cone of the two BPS rays that generated them, ensuring that even these descendants (and their own descendants) must be confined within one of $\measuredangle^\pm$ as well.
Furthermore any BPS rays within $\IR^+$ at $t=1$ are confined there also for $0\leq t < 1$, never crossing into $\measuredangle^\pm$, we provide a direct derivation in Appendix for $|\gamma|\leq 8$ and indicate how to check the statement to arbitrary order.
These facts imply that $(\ref{eq:BPS-rays-pm-virtual})$ obtained for the virtual stability condition at $t=1$ must coincide with the parts of the spectrum generator $(\ref{eq:U-split})$ for the physical stability conditions at $t=0$.

\subsection*{Computing $\IU(\IR^+)$ at $t=0$}

\begin{figure}[h!]
\begin{center}
\begin{subfigure}[b]{0.4\textwidth}
      \includegraphics[width=\textwidth]{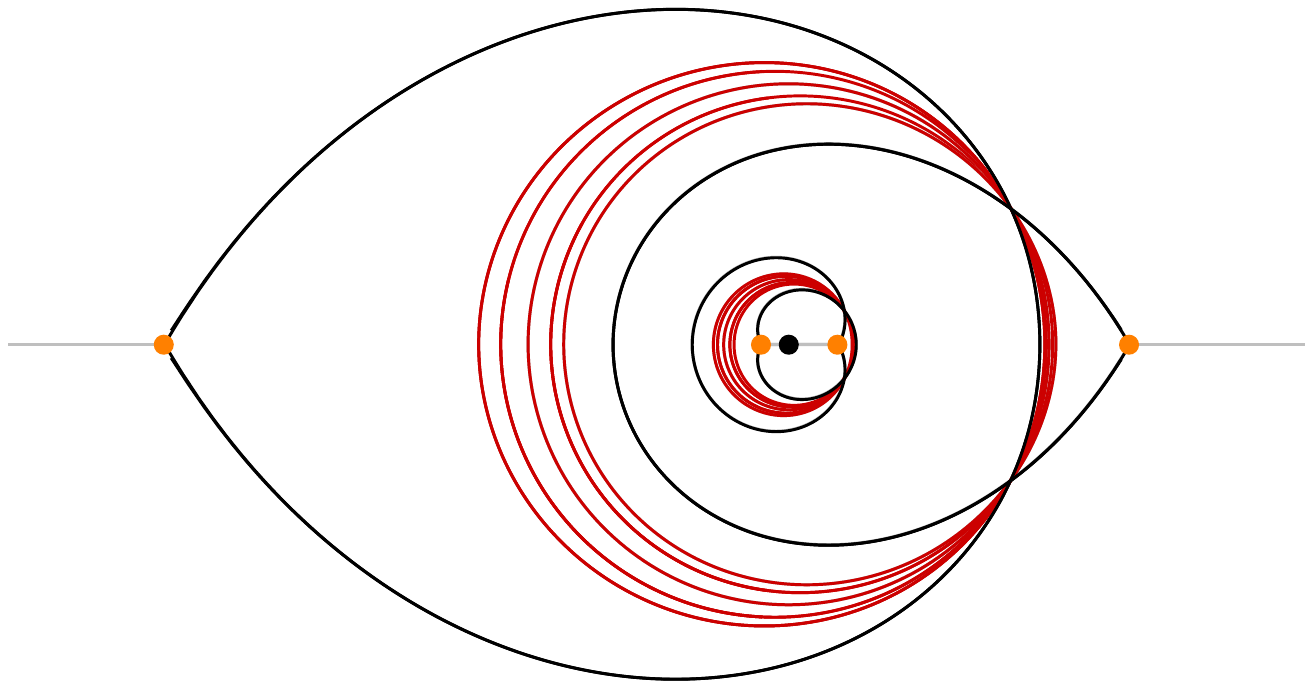}
      \caption{}
      \label{fig:theta-0-network}
    \end{subfigure}
    \begin{subfigure}[b]{0.4\textwidth}
      \includegraphics[width=\textwidth]{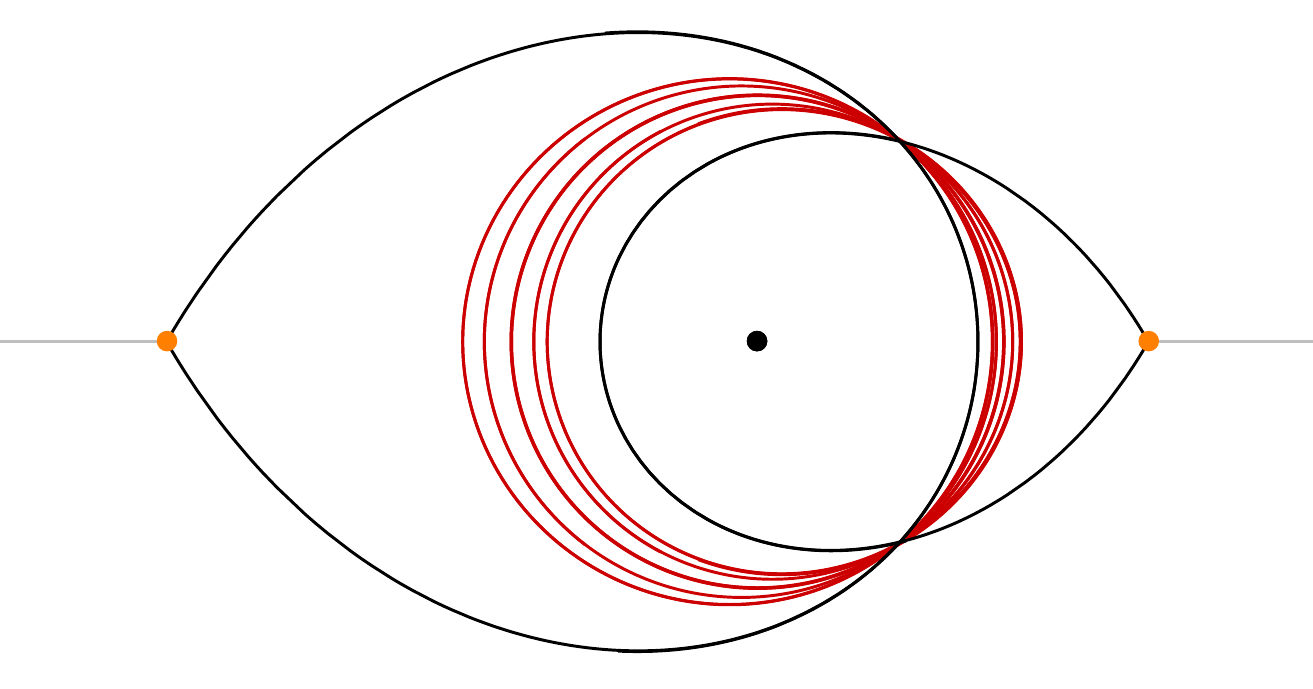}
      \caption{}
      \label{fig:C3modZ2-BPS-graph}
    \end{subfigure}
    \caption{(a) is the exponential network of local $\IF_0$ at $\vartheta=0$ for $Q_b=-1, Q_f=2$.  (b) is the exponential BPS graph for $\CO(0)\oplus \CO(-2)\to \IP^1$ in quadratic choice of framing.}
\label{fig:comparison}
\end{center}
\end{figure}

What is left out by the above analysis is to determine the part of $\IU$ corresponding to the accumulation ray.
This can be actually obtained quite easily, by plotting the exponential network at $Q_b=-1, Q_f=2$ (corresponding to $t=0$) for $\vartheta=0$, see Figure \ref{fig:theta-0-network}.

To determine the BPS spectrum encoded by the saddle one may use the machinery of \cite{Eager:2016yxd, Banerjee:2018syt, Banerjee:2019apt, Banerjee:2020moh}.
But in this case one can take a shortcut. Note that saddles are divided into two disconnected parts. 
Each set has the same topology as the exponential BPS graph of $\CO(0)\oplus \CO(-2)\to \IP^1$, shown in Figure \ref{fig:C3modZ2-BPS-graph}.\footnote{More precisely, this is the exponential network at $\vartheta=0$ for $1 + y + x y + Q y^2=0$ with $Q=6$. It is equivalent (up to framing, which has no effect on BPS states) to the curve studied in \cite[Equation (4.3)]{Banerjee:2019apt}.}
Recall that (exponential) BPS graphs encode the whole BPS spectrum of a theory \cite{Longhi:2016wtv}. 
In the case of the half-geometry, the spectrum is known to consist of $\Omega(n\, D0) = -2$ for $n\geq 1$, $\Omega(D2\-k\, D0) = -1 $ for $ k\geq 0$, and $\Omega(\overline{D2}\-k\, D0) = -1$ for $k\geq 1$ plus CPT conjugates,  see e.g. \cite[Equation (4.34)]{Banerjee:2019apt}.

To translate this result into the BPS states in $\IU(\IR^+)$ we simply have to identify $D2$ with $D2_f$ (cf. \cite[Section 5]{Banerjee:2020moh}).
Noting that $\gamma_1+\gamma_2 = D2_f$ and $\gamma_3+\gamma_4 = D0\-\overline{D2}_f$, and taking into account that Figure \ref{fig:theta-0-network} contains \emph{two} disconnected copies of the saddle in Figure  \ref{fig:C3modZ2-BPS-graph}, we arrive at the following BPS indices
\be
\begin{split}
	\Omega(n\, (\gamma_1+\gamma_2+\gamma_3+\gamma_4)) &= -4 \qquad n\geq 1\\
	\Omega(\gamma_1+\gamma_2+ k\, (\gamma_1+\gamma_2+\gamma_3+\gamma_4)) &= -2 \qquad k\geq 0\\
	\Omega(\gamma_3+\gamma_4 + k\, (\gamma_1+\gamma_2+\gamma_3+\gamma_4)) &= -2 \qquad k\geq 0
\end{split}
\ee
This prediction is quite different from \cite{Closset:2019juk}, where the states with $k>0$ and those with $n>0$ are absent.
This can be traced to the fact that \cite{Closset:2019juk} only considered stable quiver representations, whereas our result implies that $\IU$ includes also threshold states.\footnote{This interpretation was suggested to me by Michele del Zotto.\label{foot:mdz-thanks}}

To promote BPS indices to PSCs, we note that states with $\Omega=-2$ correspond to vectormultiplets (this can also be seen e.g. by the topology of their saddles), whose PSC is known to be $\Omega(\gamma;y)=y+y^{-1}$.\footnote{See \cite{Galakhov:2014xba} for a derivation of the PSC based on the topology of the saddle.}
For the states with $\Omega=-4$ one should instead note that $\gamma_1+\gamma_2+\gamma_3+\gamma_4$ is the charge of a pure $D0$ brane.
The PSC of $n \, D0$ branes in $K_{\IF_0}$ was recently argued to be $\Omega(\gamma,y) = y^{-1} (1+y^2)^2$ in \cite{Mozgovoy:2020has}.
Taking this into account completes the description of the BPS states with real central charge, leading to 
\be
\begin{split}
	\IU(\IR^+)
	& = 
	\prod_{n\geq 1}  \Phi((-y)^{-1} \hat Y_{n(\gamma_1+\gamma_2+\gamma_3+\gamma_4)})^{-1} \Phi((-y) \hat Y_{n(\gamma_1+\gamma_2+\gamma_3+\gamma_4)})^{-2}  \Phi((-y)^3 \hat Y_{n(\gamma_1+\gamma_2+\gamma_3+\gamma_4)})^{-1}\\ 
	& \times 
	\prod_{k\geq 0}
	\Phi((-y)^{-1} \hat Y_{\gamma_1 + \gamma_2 + k (\gamma_1+\gamma_2+\gamma_3+\gamma_4)})^{-1} \Phi((-y) \hat Y_{\gamma_1 + \gamma_2 + k (\gamma_1+\gamma_2+\gamma_3+\gamma_4)})^{-1} 
	\\
	& \times
	\prod_{k\geq 0}
	\Phi((-y)^{-1} \hat Y_{\gamma_3+\gamma_4+ k (\gamma_1+\gamma_2+\gamma_3+\gamma_4)})^{-1} \Phi((-y) \hat Y_{\gamma_3+\gamma_4+ k (\gamma_1+\gamma_2+\gamma_3+\gamma_4)})^{-1} 
\end{split}
\ee
As discussed earlier, all charges appearing in this expression are mutually local, ensuring no ordering ambiguities.

\section{Refined spectrum at a fiber-base symmetric point}\label{sec:sym-point-spectrum} 
The mirror geometry of $K_{\IF_0}$ has a symmetry under exchange of fiber and base moduli whenevever $Q_b = \pm Q_f$.
The point  $Q_b=-1, Q_f=1$ was studied extensively in \cite{Banerjee:2020moh}, where central charges were evaluated to be 
$Z_{\gamma_1} \approx 12.1717$,
$Z_{\gamma_2} \approx 3.15831 i$, 
$Z_{\gamma_3} \approx 27.3067$, 
$Z_{\gamma_4} \approx -3.15831 i$.
It is tedious but straightforward to compute the corresponding factorization of $\IU$, see \cite[Appendix E]{Longhi:2016wtv} for an algorithm. 
Eventually we obtain the following invariants
\be\label{eq:motivic-spectrum-sym-pt}
{\small
\begin{array}{cc}
\begin{array}[t]{|r|l|}
\hline
\gamma\qquad  & \qquad \Omega(\gamma;y)\\
\hline\hline
(n,n,n,n) & y^{3} + 2 y + y^{-1}\ \ \ (n >0)\\
\hline
(0,0,0,1) & 1\\
(0,1,0,0) & 1\\
(1,0,0,0) & 1\\
(0,0,1,0) & 1\\
\hline
(0,0,1,1) & y + y^{-1} \\
(0,1,1,0) & y + y^{-1} \\
\hline
(0,0,1,2) & 1\\
(0,0,2,1) & 1\\
(0,1,2,0) & 1\\
(0,2,1,0) & 1\\
(0,1,1,1) & y^2+ 2 + y^{-2}\\
\hline
(0,1,1,2) & y + y^{-1} \\
(0,2,1,1) & y + y^{-1} \\
(0,1,2,1) & y^3+y+y^{-1}+y^{-3}\\
\hline
(0,0,2,3) & 1\\
(0,0,3,2) & 1\\
(0,1,2,2) & y^4+ 2 y^2 +4 +2 y^{-2} + y^{-4}\\
(0,2,2,1) & y^4+ 2 y^2 +4 +2 y^{-2} + y^{-4}\\
(0,2,3,0) & 1\\
(0,3,2,0) & 1\\
(1,1,2,1) & y^2+ 2 + y^{-2}\\
(0,2,1,2) & 1\\
\hline
\end{array}
&
\begin{array}[t]{|r|l|}
\hline
\gamma\qquad  & \qquad \Omega(\gamma;y)\\
\hline\hline
(0,1,2,3) &  y^3+y+y^{-1}+y^{-3}\\
(0,1,3,2) &  y^{5}+2 y^3+4 y+4 y^{-1}+2 y^{-3}+y^{-5}\\
(0,2,3,1) &  y^{5}+2 y^3+4 y+4 y^{-1}+2 y^{-3}+y^{-5}\\
(0,3,2,1) &  y^3+y+y^{-1}+y^{-3}\\
(1,1,2,2) &  y+y^{-1} \\
(1,2,2,1) &  y+y^{-1} \\
(0,2,2,2) &  y^{5}+2 y^3+4 y+4 y^{-1}+2 y^{-3}+y^{-5}\\
\hline
(0,0,3,4) &  1 \\
(0,0,4,3) &  1 \\
(0,1,2,4) &  1 \\
(0,1,3,3) &  y^{6}+ 2 y^{4} + 4 y^2 +6 +4 y^{-2} + 2 y^{-4} +y^{-6} \\
(0,1,4,2) &  y^4 + y^2 + 2 + y^{-2} + y^{-4} \\
(0,2,2,3) &  y^4 + 2 y^2 + 4 + 2 y^{-2} +y^{-4} \\
(0,2,3,2) &  y^{8} + 2 y^{6} + 4 y^{4} + 4 y^{2} + 5 + 4 y^{-2} + 4 y^{-4} + 2 y^{-6} + y^{-8} \\
(0,2,4,1) &  y^4 + y^2 + 2 + y^{-2} + y^{-4} \\
(0,3,2,2) &  y^4 + 2 y^2 + 4 + 2 y^{-2} +y^{-4} \\
(0,3,3,1) &   y^{6}+ 2 y^{4} + 4 y^2 +6 +4 y^{-2} + 2 y^{-4} +y^{-6} \\
(0,3,4,0) &  1 \\
(0,4,2,1) &  1 \\
(0,4,3,0) &  1 \\
(1,1,3,2) &  y^2 + 2 + y^{-2} \\
(1,2,2,2) &  y^4 + 3 y^2 + 6 + 3 y^{-2} +y^{-4} \\
(1,2,3,1) &  y^2 + 2 + y^{-2}\\
\hline
\end{array}
\end{array}
}
\ee
where $(n_1, n_2, n_3, n_4)$ is the shorthand for $\sum_{i=1}^4 n_i \gamma_i$. The full spectrum is infinite, this list includes all states up to $|\gamma|\leq 7$.
Upon specialization $y\to -1$, the spectrum (\ref{eq:motivic-spectrum-sym-pt}) recovers the unrefined spectrum obtained in \cite{Banerjee:2020moh} up to degree $|\gamma|\leq 6$, and predicts several new states for $|\gamma|>6$.

\section{Discussion}

In this note we provided an exact expression for the motivic wall-crossing invariant of Kontsevich and Soibelman for the BPS spectrum of M-theory on $K_{\IF_0}$.
This operator encodes the spectrum of BPS states for any generic choice of K\"ahler moduli. In the language of gauge theory,  this is the motivic spectrum generator for BPS monopole strings and instanton particles of 5d $\CN=1$ $SU(2)$ Yang-Mills theory on $S^1\times \IR^4$.
From the viewpoint of geometry, it encodes the spectrum of rank-zero (generalized) Donaldson-Thomas invariants for $K_{\IF_0}$.

The derivation is based on data on BPS states at two different points in the moduli space of stability conditions, closely analogous to those studied in \cite[Sec 5.3]{Banerjee:2020moh} and \cite[Sec 7.2]{Closset:2019juk}.
Let us comment on how the expression $\IU$ obtained here compares with these works.

The main difference with \cite[Sec 7.2]{Closset:2019juk} lies in the factor $\IU(\IR^+)$, where we find additional infinite towers of states $\Omega(\gamma_1+\gamma_2+k\gamma_{D0};y)=\Omega(\gamma_3+\gamma_4+k\gamma_{D0};y) = y+y^{-1}$ for all $k\geq 0$, as well as  additional states $\Omega(n \gamma_{D0};y) = y^{-1}(1+y^2)^2$ for all $n\geq 1$. 
Here $\gamma_{D0} = \gamma_1+\gamma_2+\gamma_3+\gamma_4$ is the charge of a $D0$-brane, while $\gamma_1+\gamma_2$ is the charge of $D2_f$ and $\gamma_3+\gamma_4$ the charge of $D0\-\overline{D2}_f$. 
A resolution of this apparent discrepancy should be that the authors of \cite{Closset:2019juk} only studied stable quiver representations, while our results imply that $\IU$ should also include contributions from threshold states such as $D2\-D0$.${}^{\ref{foot:mdz-thanks}}$
Physically, these additional states can be expected, since they correspond to Kaluza-Klein modes of an $M2$ brane (and CPT conjugate) wrapping the fiber $\IP^1$ \cite{Gopakumar:1998ii, Gopakumar:1998jq}. The pure $D0$ states are also expected: they are directly observed in previous works e.g. \cite{Banerjee:2020moh}, and should not decay anywhere in moduli space, see e.g. \cite{Duan:2020qjy, Mozgovoy:2020has}. 
Both towers of states appear directly in exponential networks, see Figure \ref{fig:theta-0-network}.
However, since networks only compute the unrefined index $\Omega(n\gamma_{D0})=-4$, we adopted the motivic refinement $\Omega(n \gamma_{D0};y) = y^{-1}(1+y^2)^2$ from the recent work \cite{Mozgovoy:2020has}.\footnote{The counting of $D0$-branes of \cite{Mozgovoy:2020has} (which agrees with exponential networks in the unrefined limit), differs  from the \emph{physical} counting of \cite{Duan:2020qjy}. 
The relation between the two has been discussed in \cite{Mozgovoy:2020has}.}

Comparing with \cite{Banerjee:2020moh} we find direct agreement. We factorized $\IU$ to compute the refined BPS spectrum at a point with fiber-base symmetry, recovering and extending results from \cite{Banerjee:2020moh} in the limit of the spin fugacity $y\to -1$.
At generic $y$, our results agree with predictions from the Coulomb branch/Attractor Flow formulae of \cite{Manschot:2010qz, 
Manschot:2012rx, Manschot:2011xc, Manschot:2013sya, Alexandrov:2018iao}, 
computed with \cite{CoulombHiggs}. 
By extension our results should also agree with computations of Vafa-Witten invariants based on these formulae, recently carried out in \cite{Beaujard:2020sgs}, and confirming earlier predictions of \cite{Manschot:2011ym} for the spectrum of stable sheaves of arbitrary rank on Hirzebruch surfaces.

Having an \emph{exact} expression for the wall-crossing invariant, it would be very interesting to use it to study its relation to the 5d superconformal index \cite{Iqbal:2012xm}. Another interesting direction would be to compare with computations of Vafa-Witten generating functions based on different techniques, such as \cite{yoshioka1995betti, Denef:2007vg, Alexandrov:2020bwg, Alexandrov:2020dyy}.

\section*{Acknowledgements}
I would like to thank Sibasish Banerjee and Mauricio Romo for collaboration on related projects, and Fabrizio del Monte, Michele del Zotto and Boris Pioline for correspondence.
This work is supported by NCCR SwissMAP, funded by the Swiss National Science Foundation.

\appendix

\section*{Appendix: Real BPS rays from $\IZ_4$ symmetry}
Here we fill a gap left behind in the derivation above.
A crucial property of the path connecting the physical stability condition $t=0$ to the virtual one $t=1$, is that no BPS rays must cross the boundaries of $\measuredangle^\pm$ as one goes from $t=1$ to $t=0$. 
We already discussed the behavior of BPS rays within $\measuredangle^\pm$, what remains to be addressed is the behavior of the rays on $\IR^+$.

How may a positive-real BPS ray exit $\IR^+$ along the path? Since at $t=1$ we have $Z_{\gamma_1}=Z_{\gamma_3}=\overline{Z}_{\gamma_2}=\overline{Z}_{\gamma_4}$, the BPS rays in $\IR^+$ are those with charge $\gamma=(n_1,n_2,n_3,n_4)$ such that $n_1+n_3=n_2+n_4$. 
There are $(m+1)$ charges with $|\gamma|=2(n_1+n_3)=2m$, namely $\gamma = (m_1, m_2, m-m_1,m -m_2)$ for $ 0\leq m_1, m_2 \leq m$. We denote $\Gamma_0$ the set of these charges.

At $t=0$ we only found states with $n_1=n_2$ and $n_3=n_4$ in $\IR^+$, but principle there may be BPS rays which violate this condition. Notice that this may include charges which are \emph{not} mutually local with those claimed in $\IU(\IR^+)$, implying the potential presence a wall of marginal stability (MS).
This would not be a problem, since $\IU$ is well-defined (in fact it is unchanged) even on MS walls.

To keep this general possibility into account, we must find a way to study $\IU(\IR^+)$ that would work even on a MS wall. A similar problem was studied in \cite{Longhi:2016wtv}, where it was noticed that $\IU$ must exhibit certain discrete symmetries.\footnote{In \cite{Longhi:2016wtv} this was the symmetry group of a BPS graph, which is dual to a quiver \cite{Gabella:2017hpz}, making contact with the discussion here.}
In the case at hand we expect a $\IZ_4$ symmetry, generated by the following relabeling of charges in $\IU$ expressed as a formal series in $\hat Y_\gamma$
\be\nonumber
	\sigma_c : \gamma_1\to \gamma_2\to \gamma_3\to \gamma_4\to \gamma_1\,.
\ee
Since $\IU$ is independent of the stability condition, and only depends on a choice of half-plane, it has the same $\IZ_4$ symmetry of the BPS quiver. Another way to argue this symmetry, is to consider the opposite stability condition with $Z_1=Z_3, Z_2=Z_4$ but $\Arg Z_1 < \Arg Z_2$, which would produce a spectrum with an identical structure, but precisely the relabeling $\sigma_c$. Hence the spectrum generator would have the same form, with charges relabeled accordingly.
We also introduce $\sigma_d = \sigma_c\cdot \sigma_c$ which obviously exchanges $\gamma_1\leftrightarrow\gamma_3$, $\gamma_2\leftrightarrow\gamma_4$.

Let $\IU = \IU(\measuredangle^+)\, \IU_0\, \IU(\measuredangle^-)$, where $\IU_0$ is to be determined (and may differ in principle from $\IU(\IR^+)$ determined at $t=0$).
W.l.o.g. we introduce the formal series expansion $\IU_0 = \sum_{\gamma \in \Gamma_0} c_\gamma \hat Y_\gamma$ where $c_\gamma$ are functions of $y$ to be determined.

Notice that $\sigma_d$ is a manifest symmetry of $\IU(\measuredangle^\pm)$. Since the whole $\IU$ is invariant, it means that $\IU_0$ must also be invariant, thus implying $c_{\gamma} = c_{\sigma_d(\gamma)}$.
Taking this into account, we want to study the equations 
\be\nonumber
	\Eq_\gamma :  [\IU - \sigma_c(\IU)]_{\gamma} = 0
\ee
for all $\gamma = (n_1,\dots , n_4)$ with $n_i\geq 0$, and where $[F]_{\gamma}$ denotes the coefficient of $\hat Y_{\gamma}$.
Equations for $|\gamma|=2m$ and $|\gamma|=2m+1$ will only include coefficients $c_{\gamma}$ with $|\gamma|$, the system is highly overconstrained.
For example, to fix all coefficients in $\IU_0$ with $|\gamma|=2$ one only needs to consider the following subset of equations $\Eq_\gamma$ of levels $|\gamma|=2,  3$:
\be\nonumber
\begin{split}
	\Eq_{\gamma_1+\gamma_2} & : c_{(0,0,1,1)}+\frac{1+y^2}{1-y^2}=c_{(0,1,1,0)}\,, 
	\\
	\Eq_{2\gamma_1+\gamma_2} & : y^{-1}\left(y^4 c_{(0,1,1,0)} - \frac{1+y^2}{1-y^2}-c_{(0,0,1,1)}\right)=0\\
\end{split}
\ee
leading to $c_{(0,0,1,1)} = -\frac{1+y^2}{1-y^2}$, $c_{(0,1,1,0)} = 0$. It is straightforward although tedious to proceed to higher orders.
Taking into account equations $\Eq_\gamma$ to $|\gamma|=9$ we find
\be\nonumber
\begin{split}
	\IU_0 
	& = 
	1 
	- \frac{1+y^2}{1-y^2}  \( {\hat Y}_{(1,1,0,0)}+  {\hat Y}_{(0,0,1,1)} \)
	\\
	&
	+\frac{\left(2 + y^2+y^4\right) y^2 }{\left(1-y^2\right)^2 \left(1+y^2\right)} \( {\hat Y}_{(0,0,2,2)} + {\hat Y}_{(2,2,0,0)} \)
	+c_{(1,1,1,1)} {\hat Y}_{(1,1,1,1)}
	\\
	& 
	-\frac{y^4 \left(1 + 2 y^2 + y^6\right) }{\left(1-y^2\right)^3 \left(1+ y^2+y^4\right)} 
	\({\hat Y}_{(3,3,0,0)} + {\hat Y}_{(0,0,3,3)}\)
	\\
	& 
	-
	\frac{\left(1-y^2\right) \left(1-y^4\right) c_{(1,1,1,1)}-\left(2+3 y^2-y^4\right) y^2}{\left(1-y^2\right)^3}	\( {\hat Y}_{(1,1,2,2)}  + {\hat Y}_{(2,2,1,1)} \)
	\\
	& +
	\frac{y^8 \left(2+ 2 y^2 + 5 y^4 + 3 y^6+ 2 y^8 + y^{10} +  y^{12}\right) }{\left(y^2-1\right)^4 \left(y^2+1\right)^2 \left(1 + y^2 + 2 y^4 + y^6 + y^8\right)} \( {\hat Y}_{(0,0,4,4)} + {\hat Y}_{(4,4,0,0)}\)
	\\
	& +
	\frac{y^2 \left(
	2-y^2-3 y^6+y^8 + y^{12}
	\right) c_{(1,1,1,1)}+
	\left(
		1 -3 y^2 -3 y^6 + y^8
	\right) \left(1 + y^2\right)^3}{\left(1-y^2\right)^4 \left(1+y^2\right) \left(1 + y^2+y^4\right)} 
	\( {\hat Y}_{(1,1,3,3)}  + {\hat Y}_{(3,3,1,1)} \)
	\\
	& + c_{(2,2,2,2)} {\hat Y}_{(2,2,2,2)}
	+ \CO({\hat Y}_{|\gamma|=10})\,.
\end{split}
\ee
This expression shows that $\IU_0$ agrees with $\IU(\IR^+)$, fixing terms in the 3rd and 4th line  of (\ref{eq:U-expression}) up to order $k=1$, excluding the  presence of any other BPS rays on $\IR^+$ up to $|\gamma|\leq 9$. 
Evidence for the presence of all higher $k$ states includes their appearance in Figure \ref{fig:theta-0-network}, and the agreement of the spectrum  (\ref{eq:motivic-spectrum-sym-pt}) with the one of \cite{Banerjee:2020moh}. We have carried out further tests of this agreement in the unrefined limit $y\to -1$ up to $|\gamma|=12$ finding agreement with their predictions, in addition to new states (predictions of \cite{Banerjee:2020moh} are only exhaustive up to $|\gamma|=6$).
The fact that terms $c_{(n,n,n,n)}$ cannot be fixed, is related to the fact that factors $\Phi((-y)^k\hat Y_{(n,n,n,n)})$ commute with all other factors in $\IU$ and don't participate in wall-crossing, since $(n,n,n,n)$ is in the kernel of the pairing matrix.
To pin down the pure $D0$ states one needs to resort to exponential networks
or the Coulomb branch / Attractor Flow Tree formulae.
Since these don't participate in wall-crossing, they can be computed anywhere in moduli space, and Figure \ref{fig:theta-0-network} suffices for this purpose to determine $\Omega(n D0)=-4$. For the refined index one may adopt the expression of \cite{Mozgovoy:2020has}.

\bibliography{biblio}{}
\bibliographystyle{JHEP}

\end{document}